\documentclass[prb,twocolumn,floatfix,superscriptaddress,epsf,psfig]{revtex4}
\usepackage{color}
\usepackage{graphicx}
\usepackage{amsmath, amssymb}
\usepackage{lscape}
\usepackage{rotating}
\usepackage{soul}
\usepackage{color}
\usepackage{ulem}
\usepackage{epsf}
\usepackage{ulem}
\usepackage{color}
\usepackage{multirow,bigdelim}
\usepackage{array}
\usepackage{longtable}

\definecolor{newgreen}{rgb}{0.0, 0.6, 0.0}

\begin{document}
\title{Unpinned Dirac-Fermions in Carbon-Phosphorous-Arsenic Based Ternary Monolayer}
\author{Amrendra Kumar}
\affiliation {Theory and Simulations Laboratory, Theoretical and Computational Physics Section, Raja Ramanna Centre for Advanced Technology, Indore - 452013, India}
\affiliation {Homi Bhabha National Institute, Training School Complex, Anushakti Nagar, Mumbai-400094, India}
\author{C. Kamal}
\affiliation {Theory and Simulations Laboratory, Theoretical and Computational Physics Section, Raja Ramanna Centre for Advanced Technology, Indore - 452013, India}
\affiliation {Homi Bhabha National Institute, Training School Complex, Anushakti Nagar, Mumbai-400094, India}

\begin{abstract}
Graphene-like two-dimensional monolayers show many interesting and novel properties which make them potential candidates for applications in nanotechnology. In this study, we predict energetically and dynamically stable ternary  Carbon-Phosphorous-Arsenic (CPAs$_2$) monolayers in buckled geometric structure by employing density functional theory based calculations. We consider three different symmetric configurations, namely, inversion ($i$), mirror ($m$) and rotational ($r$). Binding energies of the ternary monolayers are comparable to that of silicene and better even than those of phosphorene and arsenene. The low-energy dispersions in electronic band structure and density of states (DOS) around the Fermi level contain two contrasting features: (a) parabolic dispersion around highly symmetric $\Gamma$ point with a step function in DOS due to nearly-free-particle-like Schr\"{o}edinger-Fermions and (b) linear dispersion around highly symmetric $K$ point with linear DOS due to massless Dirac-Fermions for $i$-CPAs$_2$ monolayer. The step function in DOS is a consequence of two-dimensionality of the system in which the motion of nearly-free-particles is confined. However, a closer look at (b) reveals that the ternary monolayers possess distinct characters, namely (i) massless-gapless, (ii) slightly massive-gapped and (iii) unpinned massless-gapless Dirac-Fermions for  $i$, $m$ and $r$-CPAs$_2$ configurations respectively. Thus, the nature of states around the Fermi level depends crucially on the symmetry of systems. In addition, we probe the influence of mechanical strain on the properties of CPAs$_2$ monolayer in the above mentioned three configurations. The results indicate that the characteristic dispersions of (a) and (b) move in opposite directions in energy which leads to a metal-to-semimetal transition in  $i$ and $r$-CPAs$_2$ configurations, for a few percentages of tensile strain. On the other hand, a strain induced metal-to-semiconductor transition is observed in $m$-CPAs$_2$ configuration with a tunable energy band gap. Interestingly, unlike graphene, the Dirac cones can be unpinned from highly symmetric $K$ (and $K'$) point, but they are  restricted to move along the edges ($K$-$M'$-$K'$) of first Brillouin zone due to $C_2$ symmetry in $i$ and $r$-CPAs$_2$ configurations.
\end{abstract}

\maketitle

\newpage
\section{Introduction}
\textit{Dirac material} is a class of novel materials in which the behaviour of charge carriers lying close to the Fermi level is governed by the Dirac-like equation. This is in contrast to the ordinary materials whose properties are described by the Schr{\"o}edinger equation. The Dirac materials show many exciting physical properties\cite{dirac} which are considered to be important from both fundamental as well as application perspectives. Graphene is a well known example for two-dimensional (2D) Dirac material. In this 2D monolayer, the charge carriers behave as massless Dirac-Fermions due to the presence of peculiar linear dispersion which leads to formation of Dirac cone (DC) in energy versus momentum space (E vs ($k_x$, $k_y$)) around the highly symmetric $K$ and $K'$ points in Brillouin zone (BZ) at the Fermi level. Many interesting properties, such as Klein tunneling, anomalous half-integer quantum hall effect, etc found in graphene are associated with the DC\cite{graphene,graphene1}.

In the recent past, there has been growing interest to search for new graphene-like honeycomb 2D monolayers made up of other p-block elements which may show properties similar to that of graphene and may also provide better tunability in their properties over graphene. Over the years, several theoretical and experimental investigations on p-block atom based 2D materials, both elemental (Group III: borophene, aluminene, galliene\cite{Eboro1,Eboro2,alum1,group-b12}; Group IV: silicene, germanene, stanene, plumbene\cite{sili3,sili-ck1,sili-ck2,Esi2,Esi4, group4_1,group4_2,Ege1,Esn1, plum1}; Group V: nitrogene, phosphorene, arsenene, antimonene, bismuthene\cite{nitr1,phos1,phos2,arse1,anti1,bism1,Egroup5_1,c3-sym,unpin}, etc) and binary compounds (Group IV-IV: XY with X, Y = C, Si, Ge, Sn\cite{group3-5}; Group III-V:  XY with X = B, Al, Ga, In; Y = N, P, As, Sb\cite{group3-5}; Group IV-VI: XY with X = C, Si, Ge, Sn; Y = O, S, Se, Te\cite{group-4-6}; Group-IV-V: GeP$_3$, SnP$_3$, CAs$_3$\cite{gep3,snp3-1,ck-cas3}, etc ) have  been reported in the literature. These studies uncover quite a few interesting properties of the p-block element based 2D materials. For example, the group IV based elemental monolayers possess Dirac cones at the Fermi level with zero density of states similar to that of graphene, whereas the group V based elemental monolayers are found to be semiconducting in nature. These results suggest that the 2D monolayers from the former group belong to the class of Dirac materials. On the other hand, the group III monolayers are metals, however with a signature of DCs lying close to the Fermi level\cite{group-b12}. It is observed that many p-block elemental and most of the binary monolayers (with an exception of CAs$_3$) mentioned above turn out to be the ordinary Schr{\"o}edinger materials. Thus, it is necessary to deepen the search on novel 2D Dirac materials and identification of essential requirements like symmetry, valence orbitals of constituent elements, etc for the formation of DCs is particularly important. This will enable us to narrow down the search space as well as gain insights on characteristic building blocks forming  the Dirac cone in electronic structures of 2D materials.

While searching for new p-block 2D Dirac materials, one has to also look for features which are distinct from those of graphene, but may provide a few advantages over graphene, like material compatibility with the existing semiconducting industry, better tunability in their properties, etc. For example, though silicene - a silicon analog of graphene,  possesses a linear dispersion (and hence DCs) in electronic structure similar to that of graphene, it is possible to open and tune a band gap in silicene by applying transverse external electric field\cite{sili-ck2,sili-gap4,sili-gap3}. However, similar modification is not possible in graphene. The presence of buckling in geometric structure of silicene plays a crucial role in the above process which leads to advantageous situation for silicene over graphene\cite{sili-ck2,sili-gap4}. Moreover, silicon based nanostructures are expected to be more compatible with the industry as compared to carbon based nanostructures. It is also noted that the Dirac cones observed in graphene are exclusively pinned at highly symmetric $K$ and $K'$ points in Brillouin zone due to the presence of \textit{C$_3$} rotational symmetry in planar honeycomb lattice\cite{c3-sym}. In low-energy regime, the cones are also isotropic in momentum space. However, it is  desirable to have Dirac cones at generic momentum vectors in Brillouin zone as well as anisotropic dispersions to unravel novel functionalities of 2D Dirac materials\cite{unpin,anisotropic1,anisotropic2}. This is expected to open up newer directions in applications of Dirac materials, in particular valleytronics. Breaking some of the important symmetries like  rotational \textit{$C_3$}, spatial inversion $i$, mirror $m$, etc, through either structural distortion or chemical environment, is a possible way to realize unpinned and anisotropic Dirac cones in 2D materials. Unlike the planar honeycomb geometric structure of graphene, the other 2D monolayers, namely silicene, germanene and binary CAs$_3$ are stabilized in buckled honeycomb geometric structure. Due to the buckling in geometric structure, the latter systems have lost six fold (\textit{C$_6$}) rotational symmetry leading to a change of crystal system from hexagonal to trigonal. Still, the Dirac cones in these materials are pinned at highly symmetric $K$ and $K'$ points in Brillouin zone and isotropic in low-energy regime. This suggests that further reduction in symmetry may be necessary. However, it is important to note that retaining the Dirac cones is vital and challenging while reducing the symmetry.

With the above mentioned motivations, we study the energetic, geometric and electronic properties of graphene-like honeycomb monolayer made up of Carbon-Phosphorous-Arsenic atoms (CPAs$_2$) in planar and buckled geometric structures by employing density functional theory (DFT) based electronic structure calculations. Our calculations predict energetically and dynamically stable ternary CPAs$_2$ monolayers in buckled structure. The buckled structure with spatial inversion symmetry (denoted as $i$-CPAs$_2$) is energetically more favourable by 368 meV than the planar structure. This ternary monolayer possesses the characteristic linear dispersion which forms gapless Dirac cone close to the Fermi level at highly symmetric $K$ point. Thus, the particles lying in the DC states around the Fermi level behave like massless Dirac-Fermions which makes this monolayer a 2D Dirac material. We also observe nearly-free-particle-like (NFP) states with a parabolic dispersion around $\Gamma$ point in the BZ. In addition, we identify two other geometrical configurations ($m$ and $r$-CPAs$_2$) for the monolayer by breaking the spatial inversion symmetry. The binding energy (per atom) for these two low-symmetric configurations  is very close to that of $i$-CPAs$_2$ monolayer (with a maximum difference of 10 meV/atom). Importantly, these two honeycomb structures are also found to be dynamically stable. Interestingly, these two inversion symmetry broken configurations show distinct electronic band dispersion close to the Fermi level around highly symmetric $K$ point, namely slightly massive-gapped and unpinned massless-gapless Dirac-Fermions. The remnant symmetry of 2D monolayer after breaking the spatial inversion symmetry plays a crucial role in determining the character of the DC states around the Fermi level. Moreover, we probe the influence of mechanical strain on the properties of ternary CPAs$_2$ monolayers in the above mentioned three configurations.

Remainder of the paper is arranged in the following manner. Section II provides the details of computational methods used in the present study and it is then followed by results and discussions in Section III. In the end, the paper is concluded in Section IV.

\section{Computational Details}
The self-consistent electronic structure calculations based on DFT\cite{dft} have been performed using Quantum ESPRESSO package\cite{QE}. We use the generalized gradient approximation for exchange-correlation (XC) functional given by Perdew, Burke and Ernzerhof (PBE)\cite{pbe} along with Rappe-Rabe-Kaxiras-Joannopoulos ultrasoft pseudopotentials\cite{QE-lib}.  Semiempirical Grimme's DFT-D3 method is used to account for van der Waals correction\cite{vdw}. The kinetic energy cutoff for electronic wave functions is taken as 66 Ry. We adopt Monkhorst-Pack scheme for $k$-point sampling of Brillouin zone integrations with 31$\times$31$\times$1 respectively. The convergence criteria for energy in self-consistent-field (SCF) cycles is chosen to be 10$^{-10}$ Ry. The geometric structures are optimized by minimizing the forces on individual atoms with the criterion that the total force on each atom is below  10$^{-3}$ Ry/Bohr. In order to mimic the two-dimensional system, we employ a super cell geometry with a minimum vacuum size of 18 \AA{} in a direction perpendicular to the plane of CPAs$_2$ monolayer, so that the interaction between two adjacent unit cells in the periodic arrangement is negligible.  Phonon dispersion curves are computed using density functional perturbation theory (DFPT) as implemented in PHonon package of Quantum-ESPRESSO code, with 3$\times$3$\times$1 q-mesh for all the three configurations. The geometric structure and charge density are drawn using VESTA software\cite{vesta}.

\section{Results and Discussion}
\subsection{Geometric Structure   and Stability}

\begin{figure}[t]
\begin{center}
\includegraphics[width=0.5\textwidth]{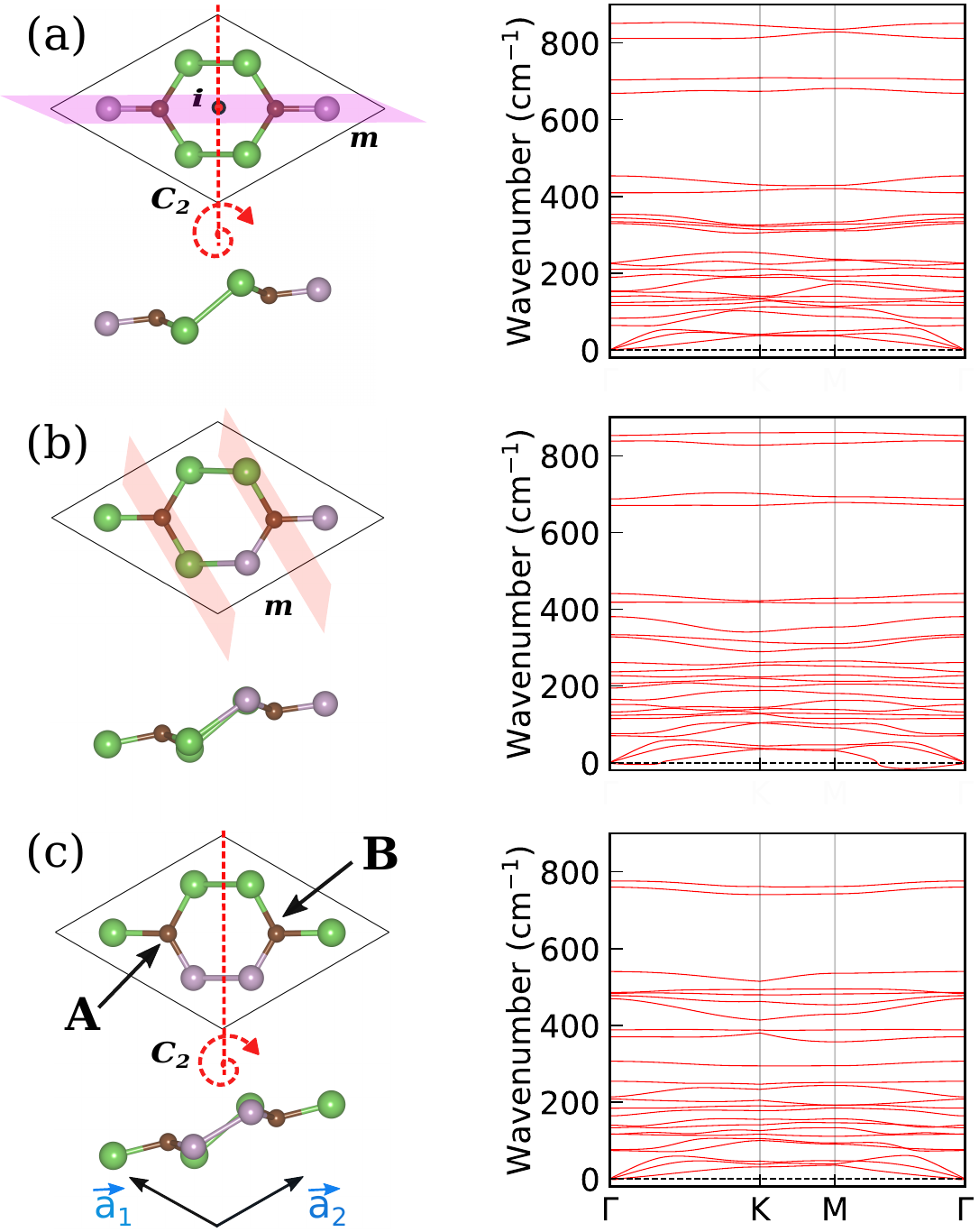}
\end{center}
\caption{Optimized buckled geometries (top and side views) and  phonon dispersions of ternary CPAs$_2$ monolayer in three symmetric configurations: (a) inversion ($i$), (b) mirror ($m$) and rotation ($r$). The symmetry operations present in the space group of each configuration are displayed. Green and purple colored balls denote arsenic and phosphorous atoms respectively. The small sized balls in brown color represent carbon atoms, which are located at sites A and B.}
\label{Fig-stuct}
\end{figure}

The results of energetics and geometrical parameters for optimized graphene-like ternary CPAs$_2$ monolayer in buckled and planar geometric structures are summarized in Table 1. Optimized geometric structures and corresponding phonon dispersions for buckled structure are given in Figure 1. The unit cell in Figure 1(a) contains two formula units of CPAs$_2$ which is equivalent to 2$\times$2$\times$1 super cell of buckled silicene monolayer. The two carbon atoms are located at (2/3, 1/3, -z) and (1/3, 2/3, +z) in the unit cell, which are denoted as sites A and B respectively. Unlike elemental silicene and binary CAs$_3$ monolayers, the ternary monolayer has lost the three-fold ($C_3$) rotational symmetry with an axis perpendicular to the sheet of monolayer since carbon atoms located at sites A and B are surrounded by inequivalent neighbours, namely one P and two As atoms. Due to this, the crystal system and thus the space group has changed from $P$-$3m1$ of trigonal to $C2/m$ of monoclinic crystal system. We note that the lattice in Figure 1(a) still retains spatial inversion symmetry which connects two carbon atoms located at sites A and B. We denote this configuration as $i$-CPAs$_2$. In addition, the monolayer possesses a two-fold rotational ($C_2$) symmetry with an axis along $\vec{a_1}$+$\vec{a_2}$ direction and a mirror plane ($m$) symmetry perpendicular to the 2D sheet which passes through the vector $\vec{a_1}$-$\vec{a_2}$. These symmetry operations are also shown in Figure 1(a).  Unlike the mirror symmetry, a $C_2$ symmetry also links the carbon atoms located at sites A and B. The values of bond distances ($d_{C-P}$, $d_{C-As}$) and angles ($\theta_{As-C-P}$, $\theta_{As-C-As}$) around carbon atoms given in Table 1 clearly indicate the rotational asymmetry with axis perpendicular to the monolayer. Importantly, the value of $\theta_{As-C-P}$ is found to be closer to 120$^\circ$ of sp$^2$  hybridization than 109.47$^\circ$ of sp$^3$ hybridization. Though the values of the bond distances $d_{C-As}$ and $d_{C-P}$ look nearly same for buckled and planar geometries, we observe large deviations in the values of bond angles $\theta_{As-C-P}$ and $\theta_{As-C-As}$.

\begin{table*}[t!]
\caption{The results of binding energy and geometrical parameters for buckled and planar CPAs$_2$ monolayers in three  symmetric configurations ($i,m,r$) obtained by DFT calculations with PBE exchange-correlation functional.}
 \begin{tabular}
{>{\centering}p{2.5cm}>{\centering}p{1.5cm}>{\centering}p{1.5cm}>{\centering}p{3cm}>{\centering}p{3cm}>{\centering}p{3cm} <{\centering}p{3cm}}\\
\hline
\hline
Parameters	&	\multicolumn{2}{c}{$i$-CPAs$_2$}  & \multicolumn{2}{c}{ $m$-CPAs$_2$}  & \multicolumn{2}{c}{$r$-CPAs$_2$}  \\  
             &    buckled &   planar   &    buckled  & planar    &    buckled &  planar  \\
\hline
\hline
Space group & $C2/m$ & $Cmmm$ & $Cm$ & $Amm2$ & $C2$ & $Amm2$ \\
E$_B$ (eV/atom) &  -3.811  & -3.443  &  -3.801 & -3.437    & -3.816 & -3.459  \\
  a (\AA{})     &  6.587 & 7.746  & 6.597  & 7.813  & 6.578 & 7.856   \\
$d_{C-As}$ (\AA{})	& 1.931	& 1.930 &  1.904, 1.965, 1.916  & 	1.925, 1.932, 1.875 & 1.933, 1.942 & 1.909, 1.925  \\
$d_{C-P}$ (\AA{})	& 1.760  & 1.734  &  1.773	&   1.749 	 & 1.749 & 1.748 \\
$\theta_{As-C-P}$ ($^\circ$)  &  118.81 &  123.72   & 114.38 & 114.57  & 119.38, 118.04 & 126.02, 122.94 \\
$\theta_{As-C-As}$ ($^\circ$)  &   111.69 &  112.55  &  115.45, 116.91 & 123.56, 118.22 & 112.48 & 111.04 \\
$\theta_{P-C-P}$ ($^\circ$)  &  -  & -   & 121.45 & 130.86 & - & - \\
\hline
\end{tabular}
\end{table*}

To probe the stability,  we calculate the binding energy and phonon dispersion spectra for this monolayer. The binding energy per atom (E$_B$) of the ternary monolayer for unit cell with eight atoms is obtained by
\begin{eqnarray}
E_B =  [E_{C_2P_2As_4}-2 E_C-2 E_P-4 E_{As}]/ 8
\end{eqnarray}
where $E_{C_2P_2As_4}$, $E_C$, $E_P$ and $E_{As}$ are the total energies of CPAs$_2$ monolayer, isolated carbon, phosphorous and arsenic atoms respectively.  Our calculations estimate that the value of E$_B$ is -3.811 eV/atom for $i$-CPAs$_2$ monolayer. From the energy stability point of view, this value is comparable to -3.96 eV/atom of silicene\cite{sili-ck1} or even better than that of other 2D monolayers such as phosphorene (-3.48 eV/atom)\cite{ec-phos},  arsenene (-2.99 eV/atom)\cite{arse1}. Moreover, we find no imaginary modes in phonon dispersions of $i$-CPAs$_2$ monolayer along high symmetry directions in the Brillouin zone (See Figure 1(a)). These results clearly suggest that $i$-CPAs$_2$ monolayer is stable in terms of both energetic as well as dynamical perspectives. We have also performed similar calculations for this ternary monolayer in planar structure. We find that this structure is energetically unfavourable by 368 meV/atom as compared to its buckled counterpart.

Spatial inversion symmetry plays a crucial role in determining the electronic structure of 2D graphene-like honeycomb structures. It is interesting to probe how the properties modify when this symmetry is broken. There are two possible ways of breaking the inversion symmetry in CPAs$_2$ monolayer by altering the environments around carbon atoms at sites A and B. This is accomplished by the rearrangements of P and As atoms. We denote these two low-symmetric structures by $m$ and $r$-CPAs$_2$ which contain only one non-trivial symmetry, namely mirror ($m$) and $C_2$ symmetry in their corresponding space groups $Cm$ and $C2$, respectively. The optimized geometries of $m$ and $r$-CPAs$_2$ monolayers in buckled structure are drawn in Figure 1(b) and (c) respectively. The results for their binding energy and  geometrical parameters are also summarized in Table 1. In case of $m$-CPAs$_2$ monolayer, three As atoms are connected to C atom at site A whereas C atom at site B has two P and one As atoms as neighbours. There is a mirror plane symmetry perpendicular to both the sheet and lattice vector $\vec{a_2}$, however it does not symmetrically link the carbon atoms at sites A and B as their environments are different. On the other hand, there exists a $C_2$ symmetry with the axis along $\vec{a_1}$+$\vec{a_2}$ in $r$-CPAs$_2$ monolayer and this symmetry connects the carbon atoms at sites A and B. We will discuss the influence of the symmetry on the electronic structure later.  Firstly, the values of binding energy for buckled $m$ and $r$-CPAs$_2$ are found to be -3.801 and  -3.816 meV/atom which are quite to close to that of $i$-CPAs$_2$ monolayer with a maximum difference of about 10 meV/atom. Secondly, the results of phonon dispersions given in Figure 1 (c) confirm the dynamical stability of $r$-CPAs$_2$  configuration as there are no imaginary modes present in the Brillouin zone. We also observe positive frequencies in the phonon dispersion spectra of  $m$-CPAs$_2$ except for the lowest phonon mode. This transverse acoustic ($ZA$) phonon mode has negative frequencies near $\Gamma$ point whose magnitude is quite small as compared to the highest frequency. Similar negative frequency has been reported in the literature\cite{group3-5} and  this may not be an instability, but occurs due to dependence of calculated frequency on the computational parameters. From Table 1, we observe large variations in bond distance $d_{C-As}$ and angle $\theta_{As-C-P}$ in buckled $m$-CPAs$_2$ monolayer as compared to those in buckled $r$-CPAs$_2$ monolayer. The value of $\theta_{As-C-P}$ is also very close to 120$^\circ$ in the latter case. Like $i$-CPAs$_2$ configuration, the values of $\theta_{As-C-P}$ (in both buckled and planar) are larger than those of $\theta_{As-C-As}$ in $r$-CPAs$_2$ configuration. However, the opposite trend is observed for $m$-CPAs$_2$ configuration. Furthermore, angle $\theta_{P-C-P}$ exists only in  $m$-CPAs$_2$ configuration and its value is larger in planar structure than that in buckled structure. Similar calculations have also been carried out for these two low-symmetric ($m$ and $r$-CPAs$_2$) configurations in planar geometric structure and the results suggest that they are energetically unfavourable by about 350 meV/atom. Since the difference in binding energies is quite large in comparison to thermal energy (25 meV) at room temperature, we exclude the ternary CPAs$_2$ in planar geometric structure for further analysis. Overall, our results clearly indicate that the ternary CPAs$_2$ monolayers favour a mixture of $sp^2$ and $sp^3$ hybridizations corresponding to buckled geometric structure rather than pure  $sp^2$ hybridization of planar geometry structure. This observation is similar to the one found in buckled monolayers such as silicene, germanene, binaries of groups IV-IV, III-V and CAs$_3$\cite{group3-5,ck-cas3}.

\subsection{Electronic Band Structure and Density of States}
\begin{figure*}[!t]
\begin{center}
\includegraphics[width=1.0\textwidth]{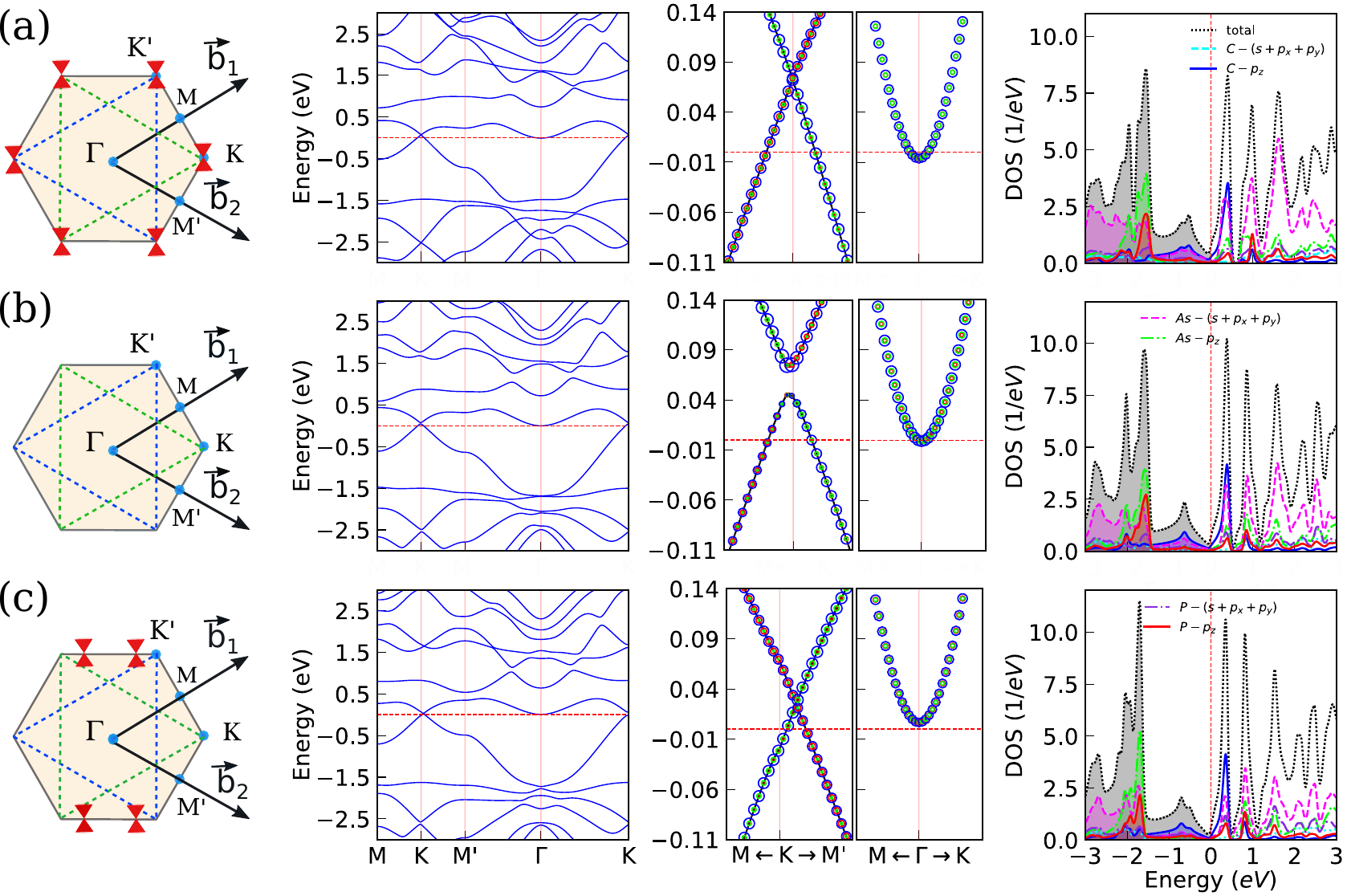}
\end{center}
\caption{ (color online) Brillouin zone (first column), electronic band structures along high symmetry paths (second column), projected bands with dispersions close to the Fermi level around highly symmetric $K$ and $\Gamma$ points (third column) and density of states (last column) for ternary CPAs$_2$ monolayer in (a) $i$, (b) $m$ and (c) $r$- configurations. Red colored cones in first column represent locations of Dirac points in Brillouin zone.}
\label{Fig-band-dos}
\end{figure*}

We calculate the electronic band structure for CPAs$_2$ monolayer in three configurations, namely $i$, $m$ and $r$ and their corresponding results are presented in (a) top, (b) middle and (c) bottom rows in Figure 2. The first column shows Brillouin zone along with the reciprocal lattice vectors and the location of Dirac points . Second and third columns of Figure 2 contain the electronic band dispersions along important highly symmetric points in the Brillouin zone and their closer look of projected bands around $K$ and $\Gamma$ points near the Fermi level respectively. Total and orbital projected density of states (DOS) are depicted in the last column. We shall first discuss the results for  $i$-CPAs$_2$ monolayer and then followed by the other two configurations.

\subsubsection{ $i$-CPAs$_2$ monolayer}
The results for  $i$-CPAs$_2$ configuration are shown in the first row of Figure 2. This monolayer is found to be  metallic in nature since there is a finite DOS at the Fermi level as a few bands cross the Fermi level. We observe two interesting features in the electronic band structures of $i$-CPAs$_2$ configuration near the Fermi level: (i) linear dispersion around highly symmetric $K$ (and $K'$) and (ii) parabolic dispersion around highly symmetric $\Gamma$ point.  The linear dispersion leads to formation of DC in three-dimensional energy versus momentum space (k$_x$, k$_y$). The crossing of these linear dispersions occurs at the so called Dirac point which lies about 72 meV above the Fermi level (E$_F$). This is in contrast to graphene, silicene and CAs$_3$ where the Dirac point lies exactly at E$_F$. Due to the presence of linear dispersion, the charge carriers in the DC states which are close to the E$_F$, behave as massless Dirac-Fermions in $i$-CPAs$_2$ monolayer. The spatial inversion symmetry, along with time-reversal, present in ternary $i$-CPAs$_2$ monolayer protects the Dirac point and thus blocks an opening of energy gap. This observation is similar to other graphene-like  honeycomb monolayers, such as silicene, germanene as well as in binary CAs$_3$\cite{sili-ck1,group3-5,ck-cas3} in buckled geometric structure. We wish to emphasize here that ternary $i$-CPAs$_2$ monolayer lacks $C_3$ symmetry and hence there is no symmetric restriction for the Dirac cone to be pinned at highly symmetric $K$ (and $K'$) point\cite{dirac}. In spite of this, the Dirac cone lies at $K$ point. This clearly indicates that the electrostatic environment around C atoms created by the neighboring As and P atoms is nearly the same. This is evident from both the valence charge distribution and Bader charge analysis presented in Figure 3(a).

On the other hand, the parabolic dispersion around $\Gamma$ point indicates the nearly-free-particle (NFP)-like character for the charge carriers. The minima of parabolic NFP-like states lies at -6 meV below the Fermi level. Thus, at the low-energy regime, the behaviour of the charge carriers in DC and NFP states is governed by relativistic Dirac-like and non-relativistic Schr\"{o}edinger-like Hamiltonians respectively. It is well known that the states around the Fermi level play an important role in determining many properties of a given material, in particular transport properties.
Hence, the properties of ternary $i$-CPAs$_2$ monolayer shall be governed by both the DC and NFP states.

To identify and probe the nature of DC and NFP states, we also perform the calculations of projected electronic band structure and DOS for $i$-CPAs$_2$ monolayer and the corresponding results are presented in third and last columns of Figure 2(a). The results of projected bands around highly symmetric $K$ and $\Gamma$ points clearly indicate that the p$_z$ orbital of carbon atoms (blue circle) contributes predominantly to both the DC and NFP states near the Fermi level and it is then followed by the p$_z$ orbitals of As (green circle) and P (red circle) atoms.  The radii of the circles in projected bands are proportional to relative contributions of the p$_z$ orbitals of constituent atoms. To get further insights, we plot the Kohn-Sham orbitals of upper and lower DC states at $K$ point as well as NFP state  at $\Gamma$ point in Figure 4(a). The two lobes of the p$_z$ orbitals on C atoms are clearly visible in both upper and lower DC states. Similar lobes are also seen in NFP states, but the sizes of lobes above and below C atoms are different.  This analysis certainly establishes an important role played by the p$_z$ orbitals of constituent atoms, in particular carbon atoms, in the formation of the Dirac cones in ternary $i$-CPAs$_2$ monolayer. This result is consistent with the previous studies on the buckled monolayers where there is a mixture of p$_z$ and other orbitals of the constituent atoms since the buckled structure do not support pure sp$^2$ hybridization as that in graphene\cite{group3-5,ck-cas3}. Note that the buckling in geometry does not break the spatial inversion symmetry present in group IV and CAs$_3$ monolayers. The results of DOS presented in the last column of Figure 2(a) show the contributions of various orbitals to the electronic states for the energy range starting from -3.0 to 3.0 eV. The contribution of p$_z$ orbitals of carbon atoms (blue curve) is dominant for the states close to the Fermi level, with minor contributions from the s, p$_x$ and p$_y$ orbitals of As and P atoms. The intense peak in the valence bands around -1.7 eV has major contributions from the p$_z$ orbitals of As and P atoms, whereas the states below -2 eV start to have stronger contributions from s, p$_x$ and p$_y$ orbitals of As and P atoms. On the other hand, we find many sharp peaks in the conduction band region. All the peaks, except the first one, have dominant contributions from s, p$_x$ and p$_y$ orbitals of As atoms and then followed by the p$_z$ orbitals of As atoms and s, p$_x$ and p$_y$ orbitals of P atoms. Importantly, the first peak occurring below 0.5 eV in conduction band region has stronger contributions from the p$_z$ orbitals of C atoms and s, p$_x$ and p$_y$ orbitals of As atoms.

Given the important role played by the p$_z$ orbitals of C atom in the formation of Dirac cones around $K$ (and $K'$) point near the Fermi level in ternary CPAs$_2$ monolayer, as observed in group IV and binary buckled CAs$_3$ monolayers, it is possible to describe the charge carriers in the DC states around the Dirac point by the following Dirac-like effective Hamiltonian in the low-energy regime\cite{oosting,wallace}.
\begin{eqnarray}
 \hat{H}= \left( {
\begin{array}{cc}
\Delta & \hbar v_F (k_x \mp ik_y)  \\
\hbar v_F (k_x \pm ik_y) & -\Delta  \\
\end{array} }
\right)
\end{eqnarray}
where $k_x$, $k_y$, $v_F$ and $\Delta$ represent the components of momentum vectors, the Fermi velocity of charge carriers near the Dirac point and the onsite energy difference between carbon atoms at sites A and B. The $+$ and $-$ signs in the Hamiltonian are used for $K$ and $K'$ points respectively. The energy eigen value for the above mentioned Hamiltonian is given by
\begin{eqnarray}
 E=\pm \sqrt{\Delta^2+ (\hbar v_F k)^2}
\end{eqnarray}
Note that there exists a spatial inversion symmetry in the ternary $i$-CPAs$_2$ monolayer since the environment around the carbon atoms at sites A and B is exactly same. This makes the onsite energy difference $\Delta = 0$. Then, the expression for energy eigen values becomes
\begin{eqnarray}
 E=\pm \hbar v_F k
\end{eqnarray}
This explains the linear dispersion around the Dirac point in the electronic band structure of $i$-CPAs$_2$ monolayer which is similar to those of graphene, other group IV and CAs$_3$ monolayers. However, unlike graphene, the Dirac point lies slightly above the Fermi level. We estimate the Fermi velocity of charge carriers in DC states to be 0.4 $\times$ 10$^6$ m/s. This value is smaller than that observed in graphene (with range 0.85 - 3 $\times$ 10$^6$),  silicene (5.27 $\times$ 10$^5$ m/s), germanene (5.09 $\times$ 10$^5$ m/s)\cite{fermi-vel} and slightly larger than that in CAs$_3$  (3.49 $\times$ 10$^5$ m/s) monolayer\cite{ck-cas3}.

On the other hand, the behaviour of charge carriers in the parabolic-like dispersion at $\Gamma$ point with a minima at -6 meV is governed by the Schr\"{o}edinger equation and the corresponding dispersion relation for nearly-free-particle like states at low-energy regime can be given as  $E=\hbar^2 k^2 / 2 m^*$, where $m^*$ is the effective mass of the charge carriers. The value of effective mass for the charge carriers in ternary $i$-CPAs$_2$ monolayer is calculated to be 0.64 m$_e$ (where m$_e$ is the mass of electron). This value lies in between the values of anisotropic effective mass (0.15, 1.2 m$_e$) in phosphorene\cite{me-phos}.
\\

\begin{figure}[!t]
\begin{center}
\includegraphics[width=0.5\textwidth]{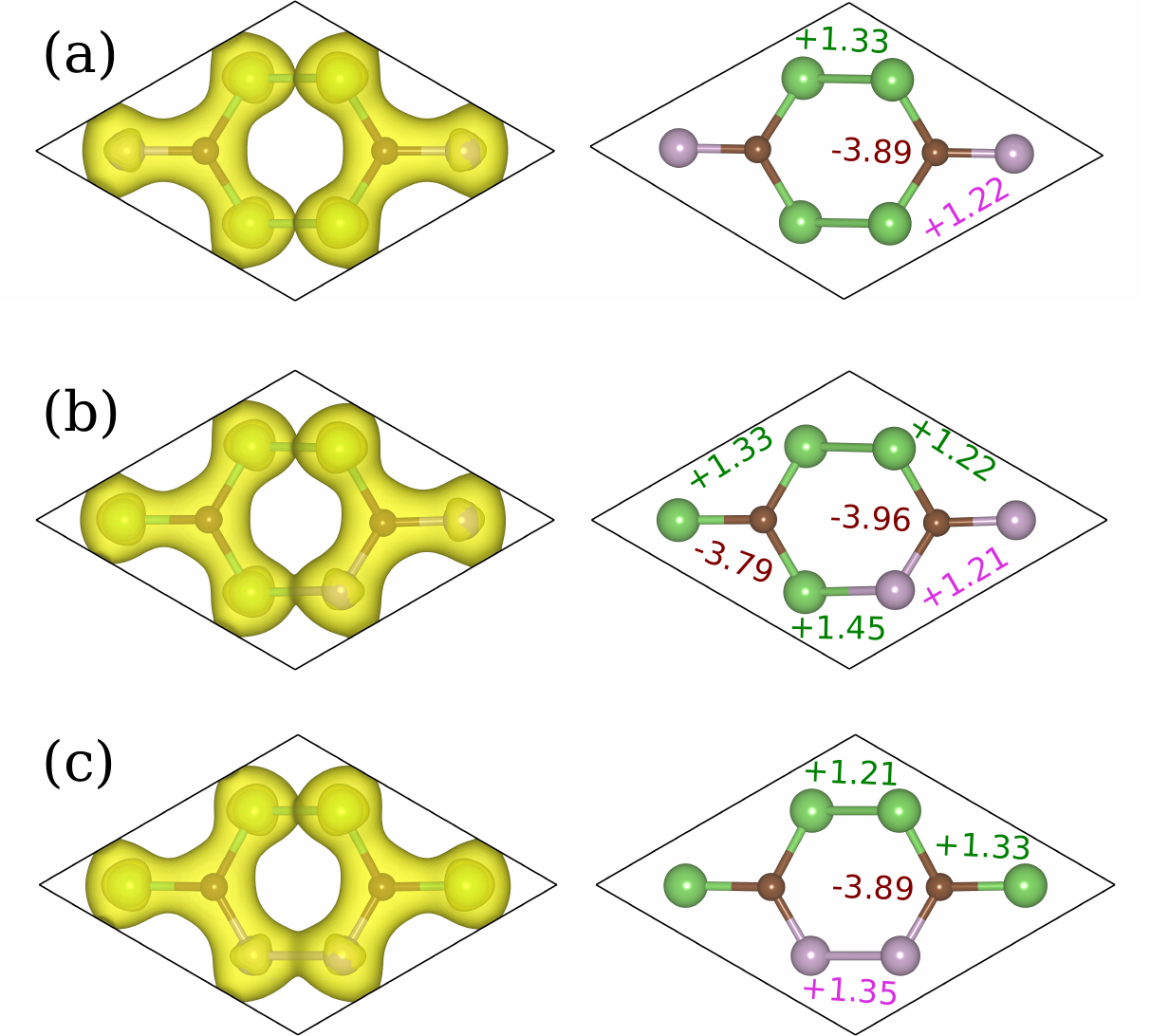}
\end{center}
\caption{ (color online) Spatial distribution of valence charge density and Bader charge on the constituent atoms in  ternary CPAs$_2$ monolayer in three symmetric configurations (a) $i$, (b) $m$ and (c) $r$. The valence charge density is plotted with an isovalue of 0.065 e/bohr$^3$}
\label{Fig-charge}
\end{figure}

\subsubsection{ $m$-CPAs$_2$ monolayer}
Now, we discuss the results for electronic structures of CPAs$_2$ monolayer in the other two configurations, namely
$m$ and $r$. These two configurations lost the spatial inversion symmetry due to the rearrangements of As and P atoms around C atoms. The electronic band structures for these two configurations are given in second column of Figure 2(b) and (c) respectively. It  is observed that the  dispersion curves for the energy range from -3.0 to 3.0 eV along the highly symmetric paths in Brillouin zone  look nearly similar to that of $i$-CPAs$_2$ monolayer.

The orbital characteristics in projected band structures and DOS of $m$-CPAs$_2$ monolayer (third and last columns of Figure 2(b)) for both the NFE-like and DC states also look similar to those of $i$-CPAs$_2$ monolayer. The Kohn-Sham orbital of NFE state  at $\Gamma$ point given in Figure 4(b) resembles that in Figure 4(a). Main reasons for these similarities are isoelectronic characters of P and As atoms and further, the difference in their electronegativities is quite small (2.19 and 2.18 eV for P and As atoms respectively\cite{en}). However, on a closer look, the dispersions around $K$ and $\Gamma$ points near the Fermi level show innate differences (see third column in Figure 2(b)). Interestingly, there is an opening of band gap causing disappearance of the Dirac point in $m$-CPAs$_2$ monolayer. This is due to difference in the environments around carbon atoms located at sites A and B as one carbon atom is surrounded by three As atoms whereas the other carbon atom has one As and two P atoms as neighbours. Note that this monolayer has the mirror plane symmetry which is perpendicular to both the sheet and $\vec{a_2}$ axis (See Figure 1(b)). However, this symmetry does not connect these two carbon atoms. Thus, as compared to $i$-CPAs$_2$ configuration, the breaking of inversion symmetry in this configuration (due to the rearrangements of P and As atoms) induces a finite value for the onsite energy difference. We have obtained the value of $\Delta=$ 16 meV by fitting the DFT based electronic band dispersions using equation (3). The small value of onsite energy difference indicates that though the carbon atoms at sites A and B have different near neighbours, the electrostatic environments provided by the neighboring P and As atoms are nearly the same. This is evident from Figure 3(b) that there are no visible differences in the valence charge distribution surrounding C atoms at sites A and B as compared to Figure 3(a) of $i$-CPAs$_2$ configuration. We also calculate Bader charges on the constituent atoms whose results are given in second column of Figure 3. This analysis also suggests that the net charges on P and As atoms differ only by small amounts. This is consistent with the fact that the electronegativity of these two atoms are nearly same. In addition, we plot the Kohn-Sham orbitals of minima and maxima of DC states around $K$ point in first and second columns of Figure 4(b) respectively.  Unlike the case of $i$-CPAs$_2$ monolayer, there is a clear asymmetrical distribution of the p$_z$ orbitals of carbon atoms at sites A and B. This happens for both the DC states  at the Dirac point leading to lifting of their degeneracy. The calculated energy gap is  32 meV (equal to twice of $\Delta$) which is slightly higher than the thermal energy at the room temperature. Fitting the energy-momentum relation using equation (3) also provides information about the Fermi velocity. We find that the slopes of upper and lower dispersion curves around $K$ point (See third column of Figure 2(b)) are slightly different which leads to dissimilar values for their Fermi velocities (the corresponding values are 0.38 $\times$ 10$^6$ and 0.46 $\times$ 10$^6$ m/s). Apart from this asymmetry, we also observe that the minimum and maximum of dispersion curves have shifted slightly away from highly symmetric point $K$.
On the other hand, the parabolic dispersion around $\Gamma$ point corresponding to the NFP-like states in $m$-CPAs$_2$ appears nearly similar to that in $i$-CPAs$_2$ monolayer with a slight upward shift (minimum lies at -1 meV). The effective mass for the charge carriers in this dispersion is calculated to be 0.68 m$_e$ which is slightly higher than that in $i$-CPAs$_2$ configuration.

\begin{figure}[!t]
\begin{center}
\includegraphics[width=0.5\textwidth]{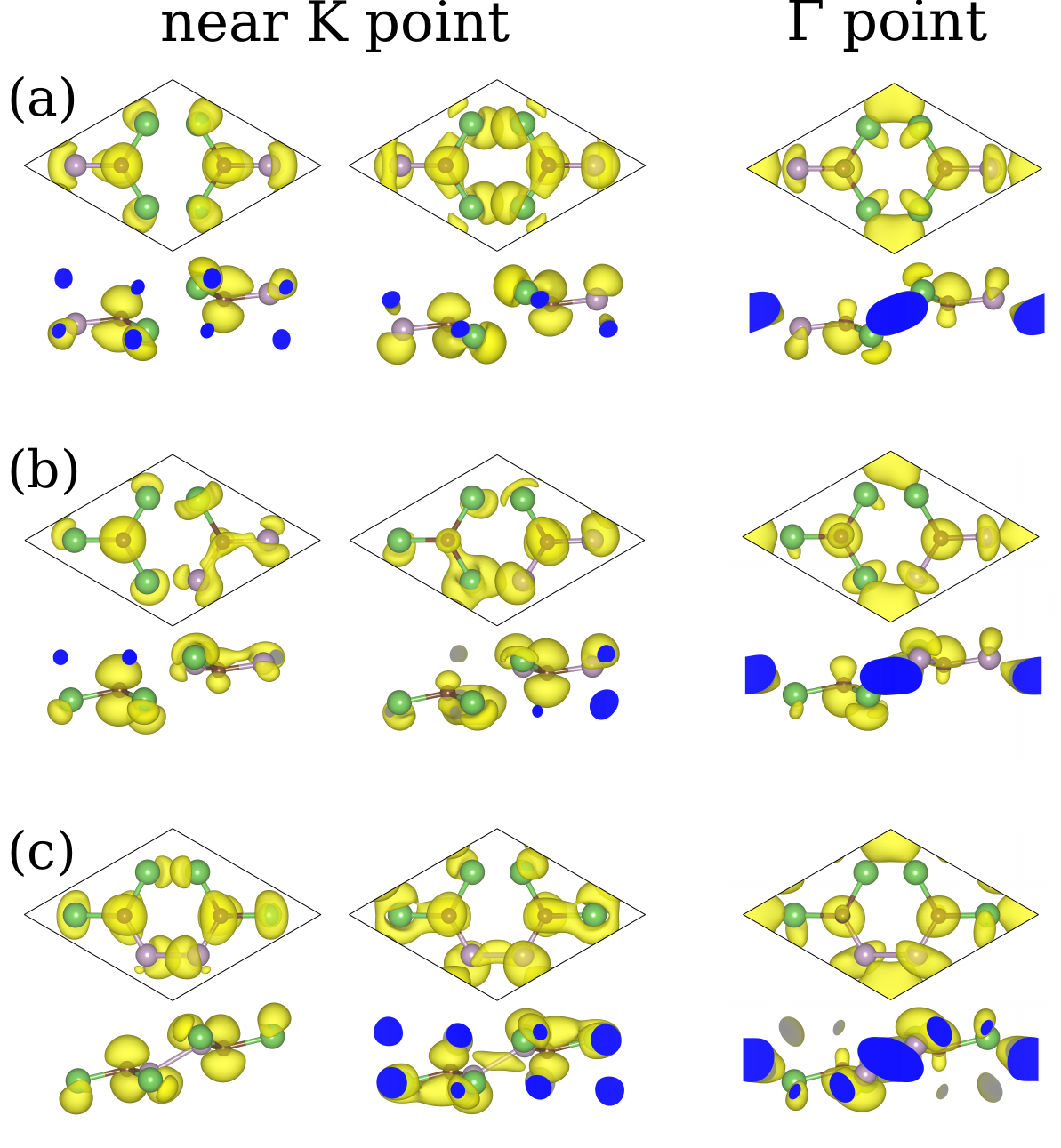}
\end{center}
\caption{ (color online) Spatial distribution of Kohn-Sham states at the minimum of upper and maximum of lower Dirac cone states near $K$ point as well as at minimum of NFP-like states around $\Gamma$ in three symmetric configurations (a) $i$, (b) $m$ and (c) $r$. }
\label{Fig-orbital}
\end{figure}

\subsubsection{ $r$-CPAs$_2$ monolayer}

In this subsubsection, we discuss the results of electronic band structure and DOS for $r$-CPAs$_2$ monolayer presented in Figure 2(c) and then compare them with those of $i$-CPAs$_2$ monolayer (Figure 2(a)). Similar to the other two configurations, $r$-CPAs$_2$ monolayer also shows metallic character.
In spite of inversion symmetry breaking, the electronic band structure of $r$-CPAs$_2$ monolayer possesses the linear dispersions and hence the Dirac cone near the Fermi level. In this configuration, the Dirac point lies much closer (at 25 meV) to the Fermi level as compared to that in $i$-CPAs$_2$ monolayer. It is important to note that, in contrast to $i$-CPAs$_2$ configuration, the Dirac point is unpinned from the highly symmetric $K$ point in  $r$-CPAs$_2$ configuration. Like in previous two configurations, we observe a NFP-like parabolic band around $\Gamma$ point in $r$-CPAs$_2$ monolayer. However,  the minimum of NFP states lies at 7 meV above the Fermi level and hence, the contribution to electronic states exactly at the Fermi level is only due to the DC states. The effective mass of the nearly-free particle is found to be 0.71 m$_e$ which is slightly higher than those of the other two configurations. Hence the mobility of charge carriers is expected to be slightly lower as compared to the previous two configurations. Analysis of the projected electronic band structures and DOS for $r$-CPAs$_2$ monolayer yields information about the  contributions of various orbitals  to the DC and NFE-like states. We find that the results are quite similar to those found in $i$-CPAs$_2$ monolayer. From Figures 3 and 4, we find that the spatial distributions of valence charge density, the NFE-like state at $\Gamma$ point, the degenerate upper and lower DC states at the Dirac point as well as the Bader charges on constituent atoms for $r$ and $i$-CPAs$_2$ configurations appear almost similar. Overall, we conclude that in this configuration also, the p$_z$ orbitals of C atoms play an important role in determining the dispersion relations around the Fermi level.

Most important difference is that the Dirac cone is unpinned from highly symmetric $K$ (and $K'$) point and has moved towards another highly symmetric point $M'$ (and $M$). To understand the unpinning of Dirac cones from the highly symmetric $K$ point, we carry out the following symmetry analysis. As mentioned in Section III.A, the space group of this monolayer only contains a non-trivial $C_2$ symmetry. This symmetry connects the two carbon atoms located at sites A and B in the unit cell. The $C_2$ symmetry along with  time-reversal symmetry preserve the linear dispersions and hence the Dirac cones in $r$-CPAs$_2$ monolayer. The pinning of Dirac cones at highly symmetric $K$ (and $K'$) point in honeycomb lattice is a restriction due to the $C_3$ symmetry. There is no such symmetric restriction for the Dirac cone to be pinned at highly symmetric $K$ point in $i$ and $r$-CPAs$_2$ configurations as both of them lack $C_3$ symmetry in their space groups. We note that the axis of $C_2$ symmetry passes through $\vec{a_1}$+$\vec{a_2}$ in direct and $\vec{b_1}$+$\vec{b_2}$ in reciprocal lattices and thus, the axis bisects the lattice vectors both in direct and reciprocal spaces. Consequently, the Dirac cones are restricted to move only on the edges ($K-M'-K'$) of first Brillouin zone which are parallel to the axis of $C_2$ symmetry. This applies to both $i$ and $r$-CPAs$_2$ configurations as both possess $C_2$ symmetry. However, it is important to find reasons for larger shift of the Dirac cone in the latter. The shifting of Dirac cone can be attributed to the asymmetrical geometry around C atoms in $r$-CPAs$_2$ monolayer. The carbon atoms at sites A and B have two As and one P atom as near neighbours in both the configurations. The three angles centred around carbon atoms with its neighbours in $r$-CPAs$_2$ configuration are symmetrically inequivalent whose value are 119.38, 118.04$^\circ$  for $\theta_{As-C-P}$  and 112.48$^\circ$ for $\theta_{As-C-As}$ whereas $i$-CPAs$_2$ configuration has 118.81$^\circ$ for $\theta_{As-C-P}$  and 111.69$^\circ$ for $\theta_{As-C-As}$. Presence of mirror plane symmetry in $i$-CPAs$_2$ configuration makes same value for two $\theta_{As-C-P}$ angles. Thus, the directional asymmetry is larger in $r$-CPAs$_2$ configuration as compared to that in $i$-CPAs$_2$ monolayer and this unpins the Dirac cone in the former. The asymmetry has also produced an anisotropy in the Dirac cones of $r$-CPAs$_2$ configuration. The calculated values of Fermi velocity along $M$-$K$ and $K$-$M'$ directions are 0.38 $\times$ 10$^6$ and 0.33 $\times$ 10$^6$ m/s for upper and 0.45 $\times$ 10$^6$ and 0.38 $\times$ 10$^6$ m/s for lower Dirac cones respectively.
Later, we will discuss in detail the influence of dissimilar values of $\theta_{As-C-P}$ angle on the electronic structure of $r$-CPAs$_2$ configuration when it is subjected to mechanical strain.
\\

\subsubsection{Characteristic Features in DOS}
\begin{figure}[!t]
\begin{center}
\includegraphics[width=0.5\textwidth]{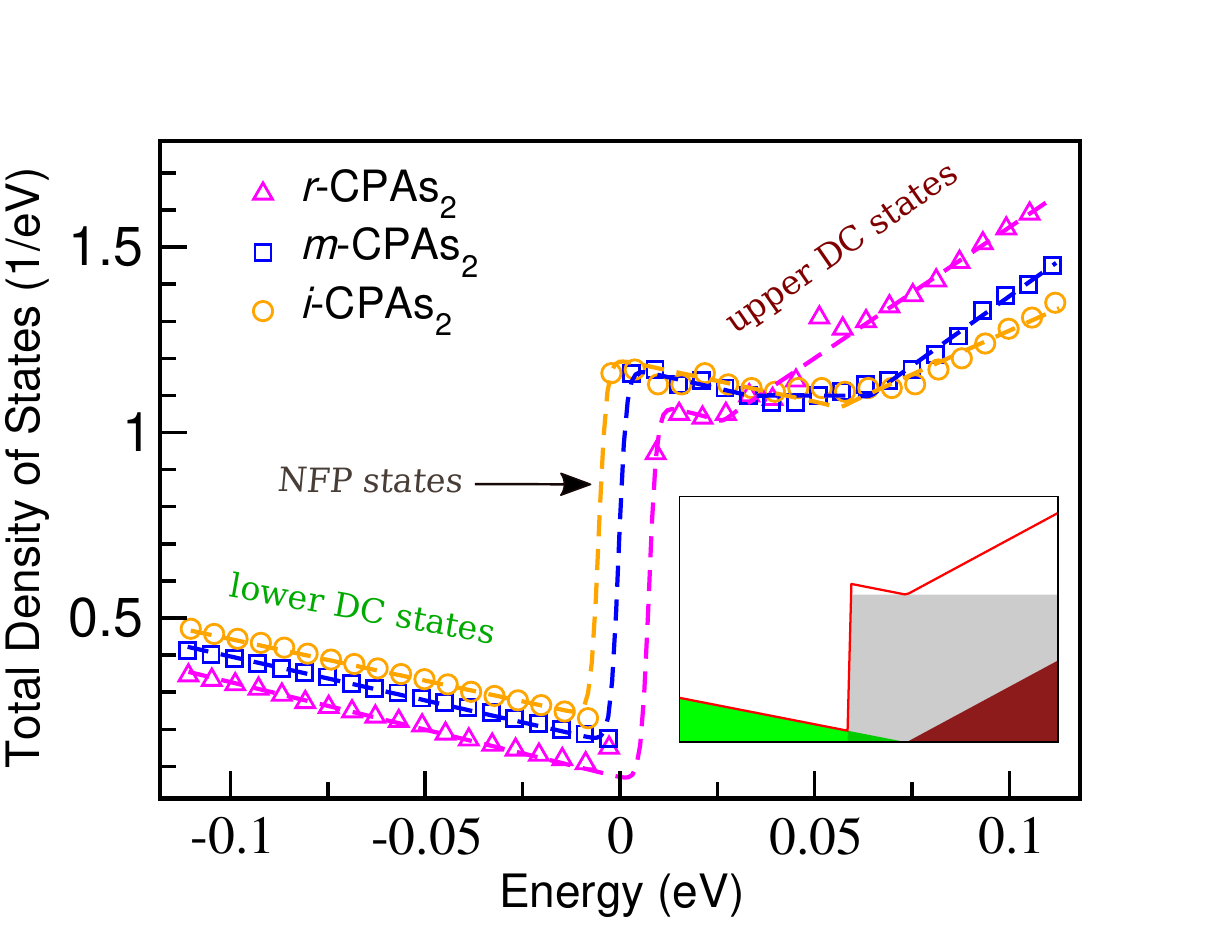}
\end{center}
\caption{ (color online) Total electronic density of states, near the Fermi level, for ternary CPAs$_2$ monolayer in three symmetric configurations ($i$, $m$, $r$). The fitted curves are shown as dashed lines.  Inset shows the schematic DOS for parabolic,  lower  and upper DC states in gray, green and brown colored shaded regions respectively. The red curve indicates the sum of these contributions. }
\label{Fig-tdos}
\end{figure}

In Figure 5, we take a closer look at the total density of states near the Fermi level for ternary CPAs$_2$ monolayer in three symmetric configurations ($i$, $m$, $r$). As we have seen in Figure 2,  there exist two types of bands near the Fermi level, namely (i) parabolic dispersion around $\Gamma$ point due to nearly-free-particle Schr\"{o}edinger-Fermion and (ii) linear dispersion due to massless-gapless (in $i$-CPAs$_2$ and $r$-CPAs$_2$)  and slightly massive-gapped (in $m$-CPAs$_2$) Dirac-Fermions. Contributions of these two characteristic dispersions to the total density of states are quite different. For Dirac-Fermions in graphene-like honeycomb lattice, the DOS is directly proportional to the modulus of energy (DOS $\propto |E-E_D|$)\cite{graphene,graphene-dos}. On the other hand, the density of states for the free particles depends strongly on dimensionality of a given system. In two-dimension, the DOS  is independent of energy and thus it shows a characteristic step function (DOS $\propto \Theta(E-E_P)$). Here $E_D$ and $E_P$ are energies of Dirac point and minimum of the parabolic dispersion respectively. Total DOS of ternary CPAs$_2$ monolayer in three configurations is sum of the two contributions mentioned above. The contributions of upper (blue color), lower (green color) Dirac cones and free particles (grey color) are schematically shown in inset of Figure 5. We have fitted the total DOS obtained from DFT calculations for CPAs$_2$ configurations with the above mentioned two functions. The fittings are quite good and the correlation coefficients are found to be above 0.99. The existence of step functions is a manifestation of two-dimensional motion of free particles (massive Schr\"{o}edinger-Fermions) present in all CPAs$_2$ monolayers. On the other hand, further analysis reveals that there are some differences in the slopes of DOS between upper and lower Dirac cones (anisotropy in Dirac cones) which can be attributed to the asymmetrical geometrical arrangements, as discussed previously, around the C atoms.

\subsection{Influence of Mechanical Strain}
\begin{figure}[!t]
\begin{center}
\includegraphics[width=0.5\textwidth]{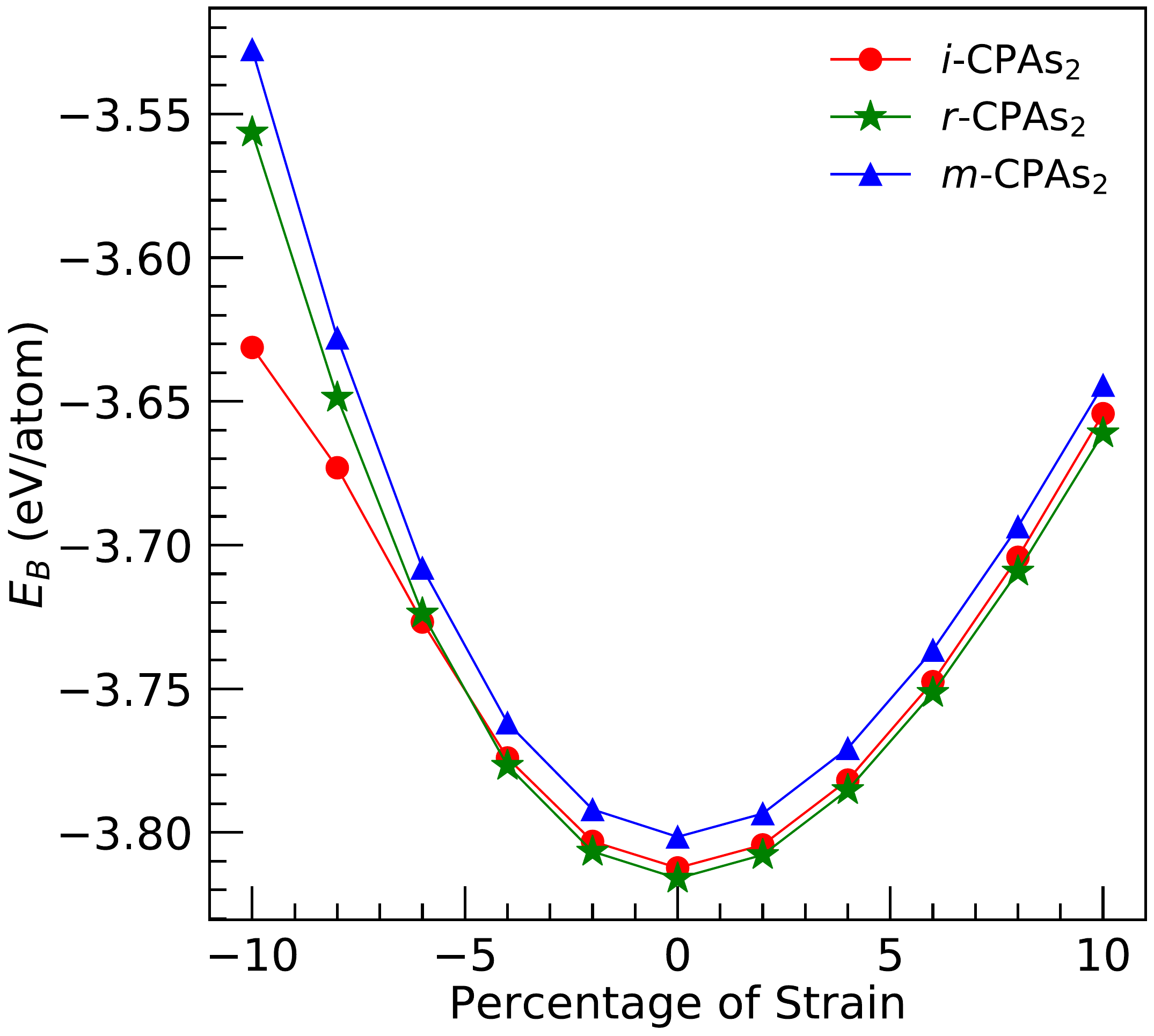}
\end{center}
\caption{ (color online) Variation in binding energy (per atom) versus the percentage of mechanical strain for ternary CPAs$_2$ monolayer in three symmetric configurations ($i$, $m$, $r$).}
\label{Fig-dos}
\end{figure}

\begin{figure*}[!t]
\begin{center}
\includegraphics[width=1.0\textwidth]{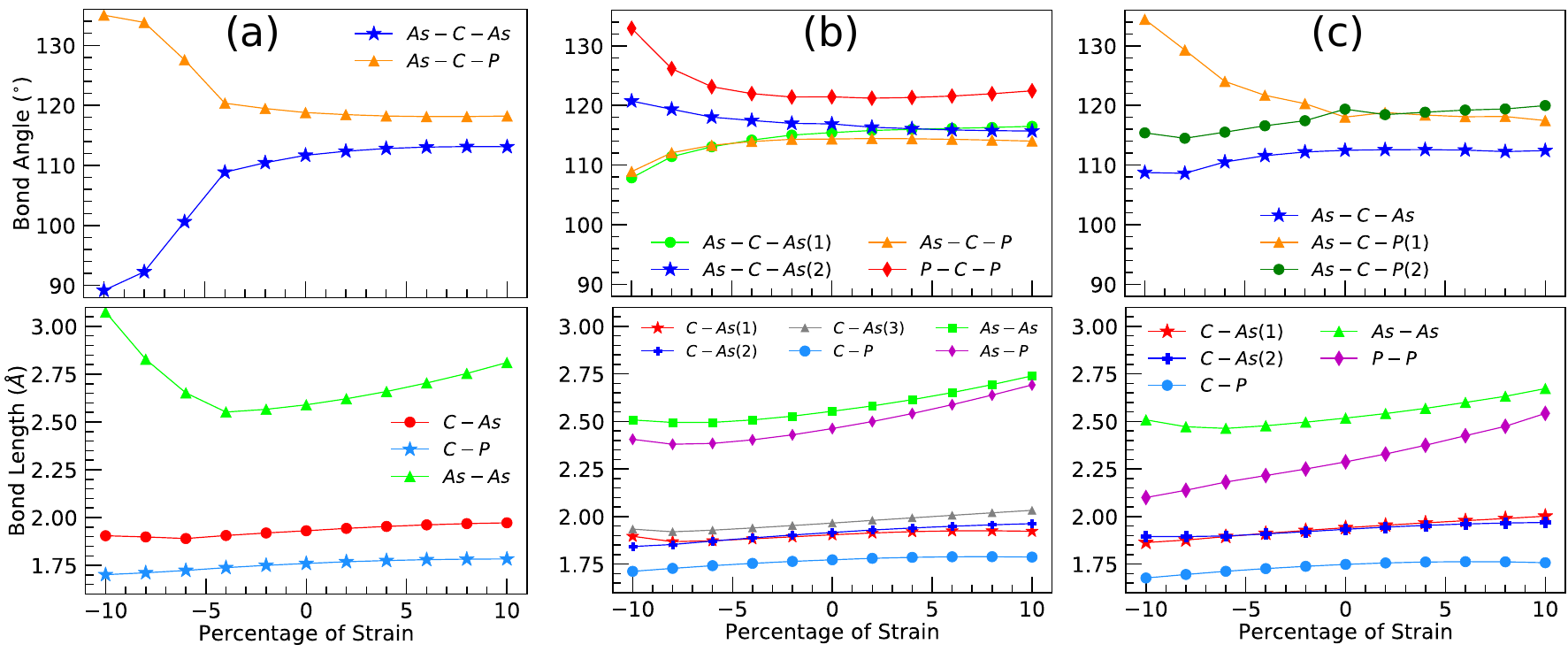}
\end{center}
\caption{ (color online) Variation of bond angles (top row) and bond lengths (bottom row) as a function of mechanical strain for ternary CPAs$_2$ monolayer in three symmetric configurations (a) $i$, (b) $m$ and (c) $r$. }
\label{fig-BL-BA}
\end{figure*}

It is interesting to probe how the properties of ternary CPAs$_2$ monolayers modify when they are subjected to mechanical strain. In the literature, there exist many investigations focused on the influences of mechanical strain on the properties of 2D materials, such as graphene, silicene, phosphorene, arsenene, group IV-VI, CAs$_3$, etc\cite{arse1, group-4-6, ck-cas3}. These studies show that there is a possibility of tuning electronic properties by applying mechanical strain. This is important for potential application of these materials to nanotechnology.

\begin{turnpage}
\begin{figure*}[!t]
 \begin{center}
\includegraphics[width=1.30\textwidth]{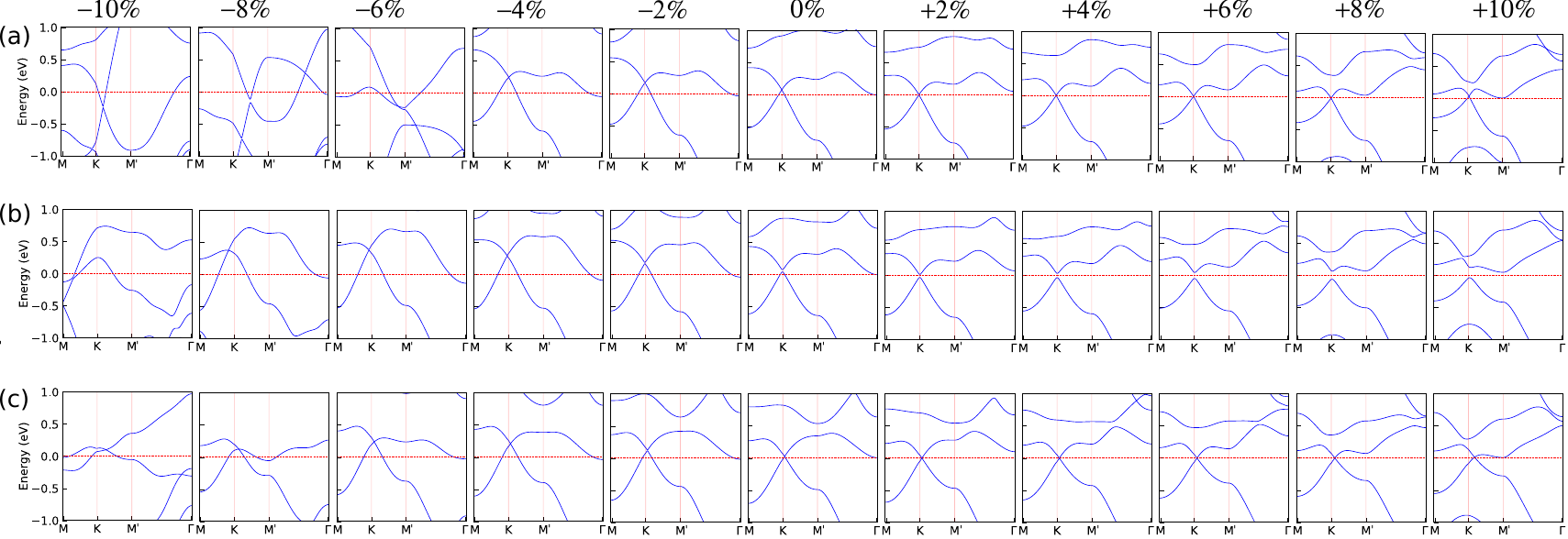}
\end{center}
\caption{ (color online) Evolution of  electronic band structures with a mechanical strain for ternary CPAs$_2$ monolayer in three symmetric configurations (a) $i$, (b) $m$ and (c) $r$. }
\label{fig-strain-band}
\end{figure*}
\end{turnpage}
 \subsubsection{On Energetics and Geometric Structure}

We investigate the energetic, geometric and electronic properties of ternary monolayers in all three configurations under the influence of biaxial strains, both compressive and tensile. Our results for the binding energy per atom as a function of strain ranging from -10 to +10\% are given in Figure 6. The negative (positive) values indicate the compressive (tensile) strains. The $E_B$ curves are nearly symmetric around the equilibrium point for a few percent of strains and start to deviate for strains beyond $-$4\% strain. Binding energy of ternary CPAs$_2$ monolayer in all three configurations shows similar trend for the tensile strain. However, we observe a deviation for $i$-CPAs$_2$ configuration for a strain beyond -6\% as compared to those of $m$ and $r$-CPAs$_2$ configurations. This deviation suggests a possibility of having drastic changes in the geometry of $i$-CPAs$_2$ configuration for compressive strain beyond -6\%. We perform a detailed analysis of the geometric structure versus strain for this configuration and the results for variations in bond lengths and angles are plotted in Figure 7(a). As we go from tensile to compressive strain, the near neighbour bond lengths ( $d_{C-As}$ and $d_{C-P}$) of carbons atoms decrease very slowly with the strain. The maximum change is less than 0.1 \AA{}, which is less than 3.5 \% of their corresponding equilibrium bond lengths, for the entire range given in Figure 7(a). Similarly, the bond angles centred around carbon atoms, both $\theta_{As-C-P}$ and $\theta_{As-C-As}$  vary slowly from +10 to -6\% of strain. Beyond -4\% of compressive strain, both the above mentioned angles modify drastically and the change is about 19$^\circ$ as we move from -4 to -10\%. In addition, the As--As bonds, which connect the two CPAs$_2$ molecules in the unit cell of $i$-CPAs$_2$ configuration, start breaking away which leads to a staggered geometric structure. The deviation in the binding energy of $i$-CPAs$_2$ configuration is due to this drastic changes in the geometry.

In cases of $m$ and $r$-CPAs$_2$ configurations, we observe from Figure 7(b) and (c) that the variations in the bond lengths are quite smooth. Note that in $m$-CPAs$_2$ configuration, carbon atom at site A (B) is attached with three As atoms (one As and Two P atoms) and there exist two inequivalent angles centered around each carbon atom due to the mirror plane symmetry. The values of angles $\theta_{As-C-As(1)}$ and $\theta_{As-C-As(2)}$ centered around C atom at site A are nearly same for tensile strain whereas they bifurcate for the compressive strain. The angles $\theta_{P-C-P}$ and $\theta_{As-C-P}$ surrounding C atom at site B also show similar trend but their values at equilibrium and during tensile strain are separated by more than 7$^\circ$. Unlike the other two configurations, the three angles around each C atom are symmetrically inequivalent, as seen in Figure 1(c) and Table 1, in $r$-CPAs$_2$ configuration. For tensile strain,  the difference between angles $\theta_{As-C-P(1)}$ and $\theta_{As-C-P(2)}$ is small, but their values are always higher than that of $\theta_{As-C-As}$. On the other hand, the former angles start deviating from each other for the compressive strain which makes the values of all three angles around C atom different. This results in stronger asymmetry around C atoms as compared to that of C atoms in other two configurations. The influence of this asymmetry on the electronic structure in $r$-CPAs$_2$ will be discussed in the following subsubsection.

\begin{figure}[!t]
\begin{center}
\includegraphics[width=0.5\textwidth]{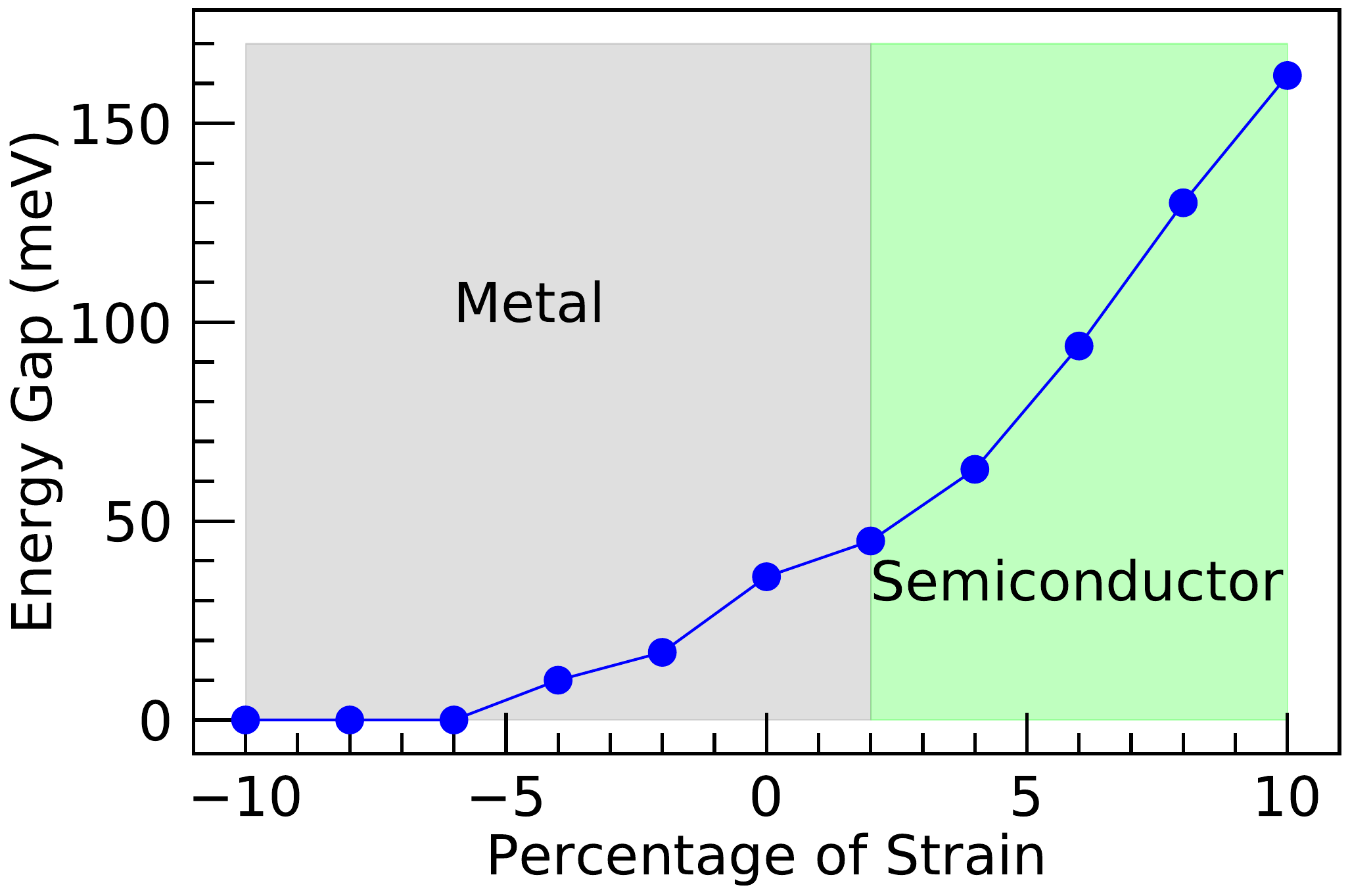}
\end{center}
\caption{ (color online) Variation of energy gap between DC states in $m$-CPAs$_2$ configuration versus the percentage of mechanical strain.}
\label{Fig-strain-gap}
\end{figure}

\subsubsection{On Electronic Band Structure}

In Figure 8, we present the evolution of electronic band structure of ternary CPAs$_2$ monolayer in all three configurations with biaxial strain, both tensile and compressive, in steps of 2\%. An important observation from Figure 8(a) and (c) is that the minima of NFP-like parabolic bands around $\Gamma$ point can be pushed above the Fermi level in $i$ and $r$-CPAs$_2$ configurations by applying tensile strain and consequently, the Dirac point, where the linear bands cross each other, has been brought exactly at the Fermi level. Importantly, the Dirac point stays exactly at the Fermi level for any further increase in the tensile strain. In this situation, the electronic states around the Fermi level are determined only by linear dispersion with Dirac-Fermionic character which is contrary to the mixed characters due to both Schr\"{o}edinger-Fermion and Dirac-Fermion in the equilibrium structures. The energy window for pure DC states is maximum at +4\% strain, for both the configurations. These modifications in the bands at $\Gamma$ and $K$ points cause $i$ and $r$-CPAs$_2$ configurations to undergo a metal-to-semimetal transition. However, for a tensile strain beyond +8\%, a band with a shallow minimum comes very close to the Fermi level at highly symmetric $M'$ point. On the contrary, the application of compressive strain brings the parabolic bands below the Fermi level and it pushes the Dirac point above the Fermi level for both $i$ and $r$-CPAs$_2$ configurations. For a few percentage of compressive strain, we observe mixed characters for the electronic states around the Fermi level as both NFP-like and DC states contribute to them. In case of $i$-CPAs$_2$ configuration, we observe major changes in the dispersion curves for a compressive strain beyond -4\%. These modifications can be attributed to the drastic changes in geometric structure as discussed in the preceding paragraphs.

Moreover, we also observe an unpinning of Dirac point from highly symmetric $K$ (and $K'$) point for both $i$ and $r$-CPAs$_2$ configurations. Unpinning of Dirac cones in ternary CPAs$_2$ monolayers is possible since the systems lack $C_3$ symmetry. However, the Dirac point is allowed to move only along the edges ( $K-M'-K'$) of first Brillouin zone due to the restriction imposed by $C_2$ symmetry, as discussed earlier. The shifting of Dirac point in momentum space for $i$-CPAs$_2$ configuration is small as compared to that in $r$-CPAs$_2$ configuration. This is mainly due to smaller asymmetric environment around C atoms at sites A and B in the former in comparison to that in the latter configuration. The presence of mirror plane in $i$-CPAs$_2$ configuration reduces the asymmetry by making two $\theta_{As-C-P}$ angles same. On the other hand, the three angles ($\theta_{As-C-P(1)}$,$\theta_{As-C-P(2)}$ and $\theta_{As-C-As}$) centred around carbon atoms in $r$-CPAs$_2$ configuration are symmetrically inequivalent. During the tensile strain, the Dirac points move away from the $K$ point due to the differences in the above mentioned angles.  Interestingly, we observe a significantly larger shift in Dirac point for compressive strain which can be attributed to the dissimilar values for the angles mentioned above.

In case of $m$-CPAs$_2$ configuration, there is an energy gap of about 32 meV around $K$ point in the equilibrium structure, but it lies above the Fermi level. It is observed from Figure 8(b) that the tensile strain shifts the parabolic band upward and brings down the energy gap at the Fermi level. This is observed even for a small percentage of strain, say +2\% which causes the system to undergo a metal-to-semiconductor transition in $m$-CPAs$_2$ configuration. As seen in previous two configurations, a shallow minima around $M'$ point comes down and this causes a direct-to-indirect band gap transition at +10\% tensile strain.
Figure 9 shows the variation of energy gap around $K$ point as a function of strain in $m$-CPAs$_2$ configuration.  We also observe that the energy gap around $K$ point can be tuned with the strain. It varies nearly parabolic fashion with respect to the tensile strain. During the compressive strain, the NFP-like parabolic band at $\Gamma$ point moves below the Fermi level and thus retaining its metallic character as that of equilibrium structure. We note that the energy gap between the DC states around $K$ point decreases with the strain and then it gets closed below -6\% of compressive strain.

Overall, the results of our calculations strongly suggest that mechanical strain can be a useful tool to tune the characters of electronic states near the Fermi level and thus the properties of ternary CPAs$_2$ monolayers.

\section{Conclusion}
We have carried out a detailed investigation on energetics, geometric and electronic properties of graphene-like ternary  CPAs$_2$ monolayer in planar and buckled geometries using density functional theory based calculations. For both geometries, we have considered three symmetric configurations, namely  $i$, $m$ and $r$ which arise due to possible rearrangements of P and As atoms. Buckled CPAs$_2$ monolayers in  these three configurations are found to be both energetically and dynamically stable. However, planar monolayer is energetically unfavourable by about 360 meV/atom as  compared to that of buckled geometry. The results of electronic structure calculations show that buckled CPAs$_2$ monolayers in these configurations are metal as they contain finite DOS at the Fermi level. Interestingly, the electronic structures, both dispersion as well as DOS, near the Fermi level display two distinct characters correspond to (a) nearly-free-particle  (NFP)-like Schr\"{o}edinger-Fermions and (b) massless Dirac-Fermions for $i$-CPAs$_2$ configuration.  These characters are identified from their symbolic parabolic (around  $\Gamma$) and linear (around $K$) dispersions as well as a step-like and a linearly dependent DOS for (a) and (b) respectively.  On the other hand, the spatial inversion symmetry broken configurations ($m$ and $r$) reveal slightly massive-gapped and massless-gapless (unpinned) Dirac-Fermions for the states around $K$ point. This indicates the crucial role played by the symmetry in determining the nature of DC states near the Fermi level whereas the nature of NFP-like states around $\Gamma$ point remain same for all the configurations. A step function in DOS is a manifestation of two-dimensional motion of nearly-free-particles in the system.

Moreover, our investigations indicate that a metal-to-semimetal transition occurs in $i$ and $r$  configurations due to the application of mechanical strain. A strain induced metal-to-semiconductor transition with a tunable energy gap is  observed in $m$-CPAs$_2$ monolayer. Finally, our results reveal that in contrast to pinned Dirac cones of graphene, Dirac cones in $i$- and $r$-CPAs$_2$ configurations can be unpinned from highly symmetric $K$ point and then moved along the edges ($K$-$M'$-$K'$) of first Brillouin zone. The present study broadens the research on two-dimensional graphene-like monolayers with tunable electronic and hence transport properties leading to a wider functionality of Dirac materials.

\section{Acknowledgment}
Authors thank the director, RRCAT and the director, Laser Group, RRCAT for constant support and encouragement. Authors also thank Dr. Aparna Chakrabarti for critical reading of the manuscript and scientific discussions and Dr. Arup Banerjee for scientific discussions. Computer Division, RRCAT is thanked for providing support in computing facility. AK thanks RRCAT, India and HBNI, India for financial support.


\end{document}